\documentclass[aps,prl,
twocolumn,
superscriptaddress,showpacs,a4paper]{revtex4-1}
\usepackage{graphicx}
\usepackage{amssymb}
\usepackage{wasysym}
\usepackage{textcomp}
\usepackage{setspace}
\usepackage[dvips,usenames]{color}
\definecolor{brown}{rgb}{0.734,0.558,0.558}
\definecolor{orange}{rgb}{1.000,0.644,0.000}

\begin{document}

\title{Anomalous field-induced growth of fluctuations in dynamics of a
  biased intruder moving in a quiescent medium}

\author{Olivier B\'enichou}
\email{benichou@lptmc.jussieu.fr}
\affiliation{Laboratoire de Physique Th{\'e}orique de la Mati{\`e}re
Condens{\'e}e (UMR CNRS 7600), Universit{\'e} Pierre et Marie Curie (Paris 6) -
4 Place Jussieu, 75252 Paris, France}
\author{Carlos Mej\'{\i}a-Monasterio}
\email{carlos.mejia@upm.es}
\affiliation{Laboratory of Physical Properties,
Technical University of Madrid, Av. Complutense s/n, 28040 Madrid, Spain}
\affiliation{Department of Mathematics and Statistics,
 P.O.  Box 68 FIN-00014, Helsinki, Finland}
\author{Gleb Oshanin}
\email{oshanin@lptmc.jussieu.fr}
\affiliation{Laboratoire de Physique Th{\'e}orique de la Mati{\`e}re
Condens{\'e}e (UMR CNRS 7600), Universit{\'e} Pierre et Marie Curie (Paris 6) -
4 Place Jussieu, 75252 Paris, France}

\date{\today}

\begin{abstract}
  We present exact results on the dynamics of a biased, by an external
  force ${\bf F}$, intruder (BI) in a two-dimensional lattice gas of
  unbiased, randomly moving hard-core particles.  Going beyond the
  usual analysis of the force-velocity relation, we study the
  probability distribution $P({\bf R}_n)$ of the BI displacement ${\bf
    R}_n$ at time {\it n}.  We show that despite the fact that the BI
  drives the gas to a non-equilibrium steady-state, $P({\bf R}_n)$
  converges to a Gaussian distribution as $n \to \infty$.  We find
  that the variance $\sigma_x^2$ of $P({\bf R}_n)$ along ${\bf F}$
  exhibits a weakly superdiffusive growth $\sigma_x^2 \sim \nu_1 \, n
  \, \ln(n)$, and a usual diffusive growth, $\sigma_y^2 \sim \nu_2 \,
  n$, in the perpendicular direction.  We determine $\nu_1$ and
  $\nu_2$ exactly for arbitrary bias, in the lowest order in the
  density of vacancies, and show that $\nu_1 \sim |{\bf F}|^2$ for
  small bias, which signifies that superdiffusive behaviour emerges
  beyond the linear-response approximation.  Monte Carlo simulations
  confirm our analytical results, and reveal a striking field-induced
  superdiffusive behavior $\sigma_x^2 \sim n^{3/2}$ for infinitely
  long 2D stripes and 3D capillaries.
\end{abstract}


\maketitle

A biased intruder (BI) traveling through a quiescent medium, composed
of bath particles which move randomly without any preferential
direction, drives the medium to a non-equilibrium steady-state with an
inhomogeneous particles' spatial distribution: the latter jam in front
of and are depleted behind the BI.  The BI can be a charge carrier
biased by an electric field or a colloid moved with an optical
tweezer. The bath particles may be colloids in a solvent or adatoms on
a solid surface.

Such microstructural changes of the medium (MCM) were experimentally
observed, e.g., in microrheological studies of the drag force on a
colloid driven through a $\lambda$-DNA solution \cite{1} or for a BI
in a monolayer of vibrated grains \cite{2}.  Brownian Dynamics
simulations have revealed the MCM for a driven colloid in a
$\lambda$-DNA solution \cite{1,3}, or for BIs in colloidal crystals
\cite{4}. Remarkably, the MCM not only enhance the drag force exerted
on the BI, but also induce effective interactions between the BIs,
when more than one BI is present \cite{14,15,16,17}.

The MCM produced by a BI were studied analytically for quiescent baths
modeled as hard-core lattice gases with symmetric simple exclusion
dynamics \cite{5,6,66,7,8,9,11}.  Despite some simplifications
(interactions are a mere hard-core, no momentum transfer and etc),
this type of modeling captures quite well many qualitative features
and reproduces a cooperative, essentially many-particle behavior
present in realistic physical systems \cite{spohn,bin}.  It is also often
amenable to analytical analysis.

It was found that in 1D the size of the inhomogeneity
grows in proportion to the traveled distance, so that the
jamming-induced contribution to the frictional drag force exhibits an
unbounded growth with time $n$. In consequence, the BI velocity $v_n \propto
n^{-1/2}$ \cite{5,6,66}, ensuring the validity of the Einstein
relation for anomalous tracer diffusion in 1D hard-core lattice gases
\cite{6,66,7,leb}.  In higher dimensions, the BI velocity approaches a
terminal value $v$ and the bath particles' distribution attains a
non-equilibrium stationary form with a jammed region in front of the
BI and a depleted region in its wake
\cite{8,9,11}.  Strikingly, behind the BI the bath particles density
approaches the mean value $\rho$ as a power-law of the distance $x$:
$1/x^{3/2}$ and $\ln(x)/x^2$ in 2D and 3D \cite{8,9,11}, which
signifies that the medium remembers the passage of the BI on large
temporal and spatial scales.

Given such an anisotropy in the bath particles' distribution, one
might be curious if the distribution $P({\bf R}_n)$ of the BI vector
displacement ${\bf R}_n=(X,Y)$ would also be asymmetric along the axis
of the applied force.  In this paper we address the question of the
asymptotic forms of the propagator and provide an exact solution
within the lattice gas model with a high bath particles density
$\rho$, (or small density $\rho_0 = 1 - \rho$ of vacancies), i.e.  for
an \textit{asymmetric} simple exclusion process (ASEP) evolving in a
2D dense sea of symmetric exclusion processes (SEPs).

Using the analytical approach developed previously by two of us in
Ref.\cite{ben}, we set out to show that, curiously enough, in the
lowest order in the density of vacancies $P({\bf R}_n)$ along the axis
of the applied force tends towards a Gaussian function as $n \to
\infty$.  Clearly, when there is no external bias so that the bath is
homogeneous, convergence to a Gaussian is quite trivial \cite{hil}.
In our case, however, given a non-equilibrium situation and essential
MCM, this result certainly cannot be expected {\em a priori}. More
striking, we will show that the variance $\sigma_x^2$ of $P({\bf
  R}_n)$ in the direction of the applied bias grows at a faster rate
due to an additional logarithmic factor, than in the direction
perpendicular to the force, which is a manifestation of a rather
counter-intuitive cooperative behavior.  Monte Carlo simulations
confirm our analytical results, and reveal a surprising superdiffusive
field-induced growth $\sigma_x^2 \sim n^{3/2}$ for infinite 2D stripes
and 3D capillaries.  On contrary, our simulations show that for
single-file geometries $\sigma_x^2$ is completely independent of the
bias.  To the best of our knowledge, these results are new.

Consider a square lattice of $L_x \times L_y$ sites ${\bf r} = (x,y)$
with integer valued components and periodic boundary conditions.  The
lattice is populated with hard-core bath particles and a single
hard-core intruder is initially at the origin. $M$ lattice sites are
vacant.  The system evolves in discrete time $n$ and particles move
randomly (subject to hard-core exclusion) by exchanging their
positions with the vacancies.  The bath particles have symmetric
hopping probabilities, i.e., given a vacancy is at an adjacent site,
any bath particles exchanges its position with the vacancy with
probability $=1/4$ independently of the direction.  On the other hand,
the intruder is subject to a constant force ${\bf F}$ oriented in the
positive $x$ direction. The normalized jump probabilities of the
({\it isolated}) BI are given, in the usual fashion (see, e.g.,
Ref.\cite{spohn}), by
 \begin{equation}
 \label{probs}
 p_{\nu} = \exp\left(\frac{\beta}{2} ({\bf F} \cdot {\bf e}_{\nu}) \right) \,/
  \sum_{\mu} \exp\left(\frac{\beta}{2} ({\bf F} \cdot {\bf e}_{\mu}) \right) \,,
 \end{equation}
 where $\beta$ is the reciprocal temperature, ${\bf e}_{\nu}$ is the
 unit vector denoting the jump direction, $\nu \in \{\pm x, \pm y\}$,
 $({\bf F} \cdot {\bf e}_{\nu})$ is a scalar product and ${\bf F} = F \, {\bf e}_{x}$ .  The sum
 with the subscript $\mu$ (the normalization constant) denotes
 summation over all possible orientations of the vector ${\bf
   e}_{\mu}$.

We turn now to the limit of small density of vacancies, $\rho_0 =
M/(L_x \times L_y) \ll 1$, and focus on the behavior in the lowest
order in $\rho_0$.  Then, it is expedient to formulate the dynamics of
the system in terms of the dynamics of vacancies.  Following
Ref.\cite{hil}, we stipulate that at each tick of the clock each
vacancy makes a step exchanging its position with a bath particle
chosen at random (with probability $=1/4$) from among its four
neighbors, in case when neither of them is the BI.  In case when one
of the neighboring particles is the BI, the situation is a bit more
complicated.  According to Ref.~\cite{ben} a correct choice of the
transition probabilities, which avoids spurious temporal trapping, is
as follows: if a vacancy is at site ${\bf R}_n + {\bf e}_{\nu}$ at
time moment $n$ and the BI occupies site ${\bf R}_n$, then it
exchanges its position with the BI with probability
$q_{-\nu} = p_{\nu}/(3/4 + p_{\nu})$,
and with the probability $= 1/(3 + 4 p_{\nu})$ with any of three
adjacent bath particles.  Note that in a complete description of the
lattice gas dynamics, these rules would have to be supplemented for
cases where two vacancies are adjacent or have common neighbors;
however, these cases contribute only to $\mathcal{O}(\rho_0^2)$ so
that we can leave the rules for such events unstated.

The steps involved in the derivation of $P({\bf R}_n)$  
are discussed in Ref.~\cite{ben}.  In the
thermodynamic limit, ($M, L_x, L_y \to \infty$ with $\rho_0 \ll1$ kept
fixed) the normalized $P({\bf R}_n)$ is given by
\begin{equation}
\label{dist}
P({\bf R}_n) \simeq \frac{1}{4 \pi^2} \int^{\pi}_{-\pi} d{\bf k} \,
\exp\left(- i \left({\bf k} \cdot {\bf R}_n\right) - \rho_0 \Omega_n({\bf k}) \right) \,,
\end{equation}
with
$\Omega_n({\bf k})$
defined via its Z-transform :
\begin{eqnarray}
\label{z}
\Omega_z({\bf k}) = \sum_{n=0}^{\infty} \Omega_n({\bf k}) \, z^n
\sim \frac{\Psi(\bf{k})}{(1-z)^2 \left(1 + \chi_z^{-1} \Psi(\bf{k})\right)} \,,
\end{eqnarray}
where $"\sim"$ denotes the exact
leading at $z \to 1^-$ ($n \gg 1$)  behavior,
$\chi_z \sim - \pi/\ln(1-z)$ 
is the generating function of the 
mean number $\chi_n$ of "new" sites visited at the $n$-th step (see, e.g., Ref. \cite{hil}),
 $\Psi({\bf k}) = - i a_0 k_x + a_1 k_x^2/2 + a_2 k_y^2/2$ and
$a_j$ are given, for arbitrary $\xi =\beta F/2$, by
\begin{equation}
\label{a0}
a_0 = \frac{\sinh(\xi)}{(2 \pi - 3) \cosh(\xi) + 1} \,, \,\, a_1 = \coth(\xi) \, a_0 \,,
\end{equation}
\begin{equation}
\label{a2}
a_2 = \frac{1}{\cosh(\xi) + 2 \pi - 3} \,.
\end{equation}

Differentiating $\Phi({\bf k}) =
\exp\left(- \rho_0 \Omega_n({\bf k})\right)$ with respect to $k_x$ and $k_y$,
we find that the BI mean velocity $v$ and the variances $\sigma_y^2$ and $\sigma_x^2$ along the $y$-  and $x$-axes obey
\begin{eqnarray}
v &\sim& \rho_0 \, a_0 \,, \label{eq:vel}\\
\sigma^2_{y} &\sim& \rho_0 \, a_2  \, n \,, \label{2}\\
\label{1}
\sigma^2_{x} &\sim& \rho_0 \, \left(a_1 +  \frac{2 a_0^2}{\pi} \, \left(H_{n+1} - 1\right)\right) \, n,
\end{eqnarray}
where $H_{n} = \sum_{k=1}^{n} k^{-1}$ is the $n$-th harmonic number.

\begin{figure}[ht]
  \centerline{\includegraphics*[width=0.5\textwidth]{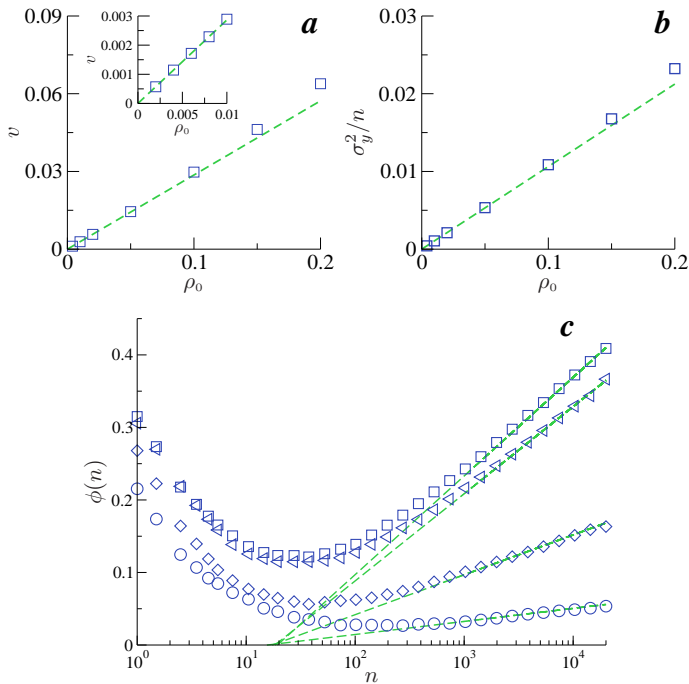}}
  \caption{(color online) Velocity $v$ (panel {\it a}) and
    $\sigma_y^2/n$ (panel {\it b}) versus $\rho_0$ for $\beta F=5$.
    The dashed lines define our theoretical predictions in
    Eqs.~(\ref{eq:vel}) and (\ref{2}).  The symbols are
    the MC simulations results. The inset in panel {\it a} is a
    zoom of the region $\rho_0 \le 0.01$. In panel {\it
      c} we plot $\phi(n)$ (see the text) vs $n$ for
    $\rho_0=0.002$, $\beta F = 100$ (${\color{blue} \square}$), $\beta
    F = 5$ (${\color{blue} \vartriangleleft}$), $\beta F = 2$
    (${\color{blue} \Diamond}$) and $\beta F = 1$ (${\color{blue}
      \fullmoon}$). The dashed curves are our predictions in
    Eq.~(\ref{11}).  }
  \label{fig-2}
\end{figure}

In Fig.~\ref{fig-2} (panels {\it a} and {\it b}) we present a
comparison of our analytical results for the mean velocity and the
variance in the $y$-direction and the results of numerical simulations
for fixed bias $\beta F = 5$ and different values of $\rho_0$. One
notices that our analytical results are in a very good agreement with
the numerical data for $\rho_0$ up to $0.2$.

Noticing next that in the large $n$ limit $H_{n+1} \sim \ln(n) +
\gamma + \mathcal{O}(1/n)$, where $\gamma \approx 0.577$ is the
Euler-Mascheroni constant, we find that asymptotically
\begin{equation}
\label{11}
\sigma^2_{x} \sim \rho_0 \, \left(a_1 + \frac{2 a_0^2}{\pi} \left(\gamma - 1\right) + \frac{2 a_0^2}{\pi} \, \ln(n)\right) \, n \,.
\end{equation}
On comparing the results in Eqs.~(\ref{2}) and (\ref{11}), we 
thus conclude that $\sigma_x^2$ grows faster  due to an
additional logarithmic factor than $\sigma_y^2$,
which shows the usual diffusive behavior.  It means that, rather counter-intuitively, the
distribution becomes progressively more broad along the $x$ direction,
compared to its behavior along the $y$ direction.  Note that, since $a_0 \sim \beta F$ for $\beta F \ll 1$, the coefficient before the superdiffusive term is proportional to $(\beta F)^2$, which signifies that the superdiffusive behavior emerges beyond (and thus can not be obtained within) the linear-response approximation.

In Fig.~\ref{fig-2} (panel {\it c}) we present an evidence for the
additional logarithmic factor in the $x$-component of the variance. To
single out this contribution, we plot here versus $\ln(n)$ the
function $\phi(n) = \sigma_x^2/n - \rho_0 (a_1 +2 a_0^2
(\gamma-1)/\pi)$, which according to Eq.~(\ref{11}) should grow in
proportion to $\ln(n)$. One observes an apparent crossover to a
logarithmic behavior which persists then over two time decades.

The skewness $\gamma_1(x)$ of $P(X) = \int dY
\, P({\bf R}_n)$
obeys
\begin{equation}
\label{skewness}
\gamma_1(x) \sim \frac{6 a_0 \left(a_0^2 + \pi a_1/\ln(n)\right)}{\left(2 a_0^2 + \pi a_1/\ln(n)\right)^{3/2}} \,
 \sqrt{\frac{\ln(n)}{\pi \, \rho_0 \, n}} \,.
\end{equation}
Note that $\gamma_1(x) > 0$ so that $P(X)$ has a
\textit{positive} skew which implies that fluctuations in the BI
position are more pronounced for $X > v \, n$ (i.e. in the region
where the bath particles jam) than for $X < v \, n$ where the bath
particles are depleted.
Next, for the kurtosis
 we get
\begin{equation}
\label{kurtosis}
\gamma_2(x) \sim \frac{6 \left(4 a_0^4 + 6 a_0^2 \pi a_1/\ln(n) + \pi^2 a_1^2/\ln^2(n)\right)}{\left(2 a_0^2 + \pi a_1/\ln(n)\right)^2} \, \frac{\ln(n)}{\pi \rho_0 \, n} \,,
\end{equation}
i.e.,
$\gamma_2(x)$ vanishes
as $n \to \infty$, and hence, $P(X)$  \textit{converges} to a Gaussian,
despite a non-equilibrium and MCM!

We find next that the skewness 
$\gamma_1(y)$
of
$P(Y) = \int dX \, P({\bf R}_n)$ vanishes. For the kurtosis of $P(Y)$
 we obtain
$\gamma_2(y) \sim 6 \ln(n)/\pi \, \rho_0 \, n$,
which implies that $\gamma_2(y)$ decays exactly in the same way as
$\gamma_2(x)$ and moreover, $\gamma_2(y)/\gamma_2(x) \to
1^-$ as $n \to \infty$.  Observe that $\gamma_2(y)$ is
\textit{independent} of the bias, while $\gamma_1(x)$ and $\gamma_2(x)$
become \textit{independent} of it when $\ln(n) \gg \pi
a_1$.

\begin{figure}[ht]
  \centerline{\includegraphics*[width=0.5\textwidth]{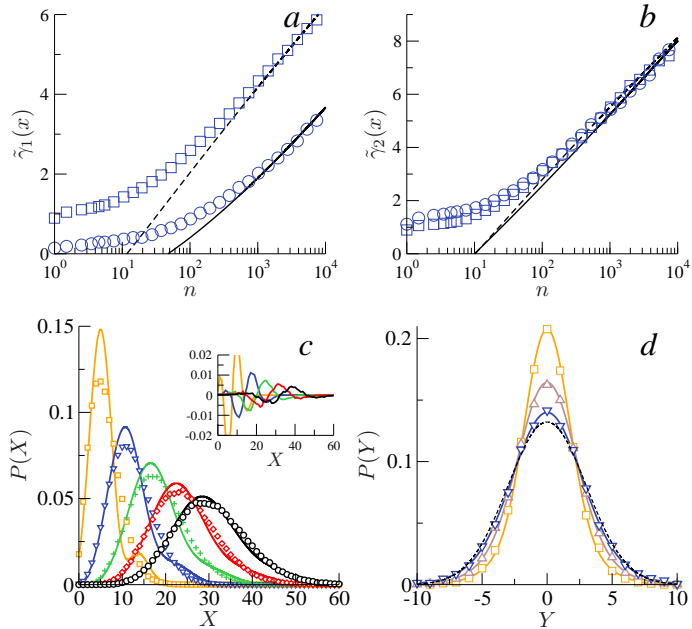}}
  \caption{(color online) Reduced skewness $\tilde{\gamma}_1(x) = n \,
    \gamma_1^2(x)$ (panel {\it a}) and the kurtosis
    $\tilde{\gamma}_2(x) = n \gamma_2(x)$ (panel {\it b}) for $\rho_0
    = 0.002$, $\beta F = 1$ (${\color{blue} \fullmoon}$) and $\beta F
    = 100$ (${\color{blue} \square}$).  Dashed and solid curves are the
    corresponding theoretical results in Eqs.~(\ref{skewness}) and
    (\ref{kurtosis}).  Panels {\it c} and {\it d} present the time
    evolution of the integrated distributions: $P(X)$ for
    $\rho_0=0.002$ and $\beta F=100$ for times $n = 10^4$
    (${\color{orange} \square}$), $2 \times 10^4$ (${\color{blue}
      \bigtriangledown}$), $3 \times 10^4$ (${\color{green} +}$), $4
    \times 10^4$ (${\color{red} \Diamond}$) and $10^5$
    (${\color{black} \fullmoon}$).  The inset shows the deviation
    $\Delta P(X)$ of our result in Eq.~(\ref{G}) from the numerical
    data.  Panel ${\it b}$: $P(Y)$ for $\rho_0=0.002$ and $\beta F=1$
    for $n = 10^4$ (${\color{orange} \square}$), $1.5 \times 10^4$
    (${\color{brown} \bigtriangleup}$) and $2 \times 10^4$
    (${\color{blue} \bigtriangledown}$).  Solid curves are our results
    in Eqs.~(\ref{G}) and (\ref{GG}).  The black dashed line defines
    the asymptotic Gaussian distribution for $n = 2 \times 10^4$.}
  \label{fig-5}
\end{figure}

In Fig.~(\ref{fig-5}) (panels {\it a} and {\it b}) we compare our
analytical predictions for $\gamma_1(x)$ and $\gamma_2(x)$ in
Eqs.~(\ref{skewness}) and (\ref{kurtosis}) against the results of
numerical simulations (symbols). To single out the logarithmic
factors, we plot here the functions $\tilde{\gamma}_1(x) = n \,
\gamma_1^2(x)$ and $\tilde{\gamma}_2(x) = n\gamma_2(x)$ vs
$\ln(n)$. Note that the kurtosis approaches the asymptotic prediction
in Eq.~(\ref{kurtosis}) quite rapidly, at times of order $n \sim 10^2$
(and moreover, the $\beta F$-independent behavior is established
quite fast as well). This confirms our result that the kurtosis
vanishes as $\gamma_2(x) \propto \ln(n)/n$ and, hence, that the
distribution in the direction of the bias converges to a Gaussian.
The skewness, which decays at a slower rate, approaches the asymptotic
result in Eq.~(\ref{skewness}) a decade later and the  $\beta F$-independent  behavior
 sets in at larger times, which are not
accessible in our numerical simulations of this essentially
many-particle system.

We turn next to the large- (but finite) $n$ corrections to the
asymptotic Gaussian distribution.  Multiplying
both denominator and the nominator of Eq.~(\ref{z}) by the complex
conjugate of the denominator, introducing a small parameter
$\epsilon_x = a_1 \ln(n)/2 \pi \sigma_x^2 \propto 1/n$ ($\epsilon_y =
a_2 \ln(n)/2 \pi \sigma_y^2 \propto \ln(n)/n$) and expanding
$\Phi(k_x,k_y=0)$ ($\Phi(k_x=0,k_y)$) up to the second order in
$\epsilon_x$ ($\epsilon_y$), we get
\begin{widetext}
\begin{eqnarray}
\label{G}
P(X) &=& \frac{\exp\left( - \left(\eta_x - \overline{\eta}_x\right)^2/2\right)}{\sqrt{2 \pi \sigma_x^2}} \,
 \,
\Big\{1
+ \Big[ 3 \, \big( g - f  \big)
+ 6 \, f \, \overline{\eta}_x^2 -  \big( g + f  \big)\, \overline{\eta}_x^4
+ 2 \, \Big( 3 \, \big(g - 2 f  \big) +  \big(g + 2 f  \big) \, \overline{\eta}_x^2 \Big) \, \overline{\eta}_x \, \eta_x \nonumber\\ &-& 6 \, \Big(g - f  \, + f \, \overline{\eta}_x^2  \Big) \, \eta_x^2
- 2 \Big(g - 2 f \Big) \, \overline{\eta}_x \, \eta_x^3 + \Big(g -  f \Big) \, \eta_x^4
 \Big] \, \epsilon_x + \mathcal{O}(\epsilon_x^2 \ln(n))\Big\}\,,
 \end{eqnarray}
 \begin{eqnarray}
\label{GG}
P(Y) &=& \frac{\exp\left( - \eta_y^2/2\right)}{\sqrt{2 \pi \sigma_y^2}} \Big\{1 +
 \left[3 - 6 \eta_y^2 + \eta_y^4\right] \, \frac{\epsilon_y}{2} + \mathcal{O}(\epsilon_y^2)\Big\} \,.
\end{eqnarray}
\end{widetext}
where we have conveniently chosen $\eta_x= X/\sigma_x$ ($\eta_y=
Y/\sigma_y$), $\overline{\eta}_x = v n/\sigma_x$, $g = 1+ a_0^2
\ln(n)/\pi a_1$ and $f = \pi a_1/ 2 (\pi a_1 + 2 a_0^2 \ln(n))$.  Note
that the result in Eq.~(\ref{G}) becomes identical to that in
Eq.~(\ref{GG}) when $\beta F = 0$.

In Fig.~\ref{fig-5} (panels {\it c} and {\it d}) we compare our
Eqs.~(\ref{G}) and (\ref{GG}) against the numerical simulations
data. We observe that as time progresses, the discrepancy between
Eq.~(\ref{G}) and numerically obtained $P(X)$ gets smaller, as
evidenced by the inset in panel {\it c}. The convergence is
non-uniform so that we have a better agreement between the theory and
numerics for the values of $X$ to the left from the maximum, than to
the right from it (recall that $\gamma_1(x) > 0$).  Along the $y$-axis
(panel {\it d}), we observe a pretty good agreement between our
Eq.~(\ref{GG}) and the numerical data. Note, however, that some slight
discrepancy between a Gaussian (blue dashed line) and a Gaussian with
the correction term, Eq.~(\ref{GG}), persists even for the longest
observation time, $n = 2 \times 10^4$.

\begin{figure}[!b]
  \centerline{\includegraphics*[width=0.5\textwidth]{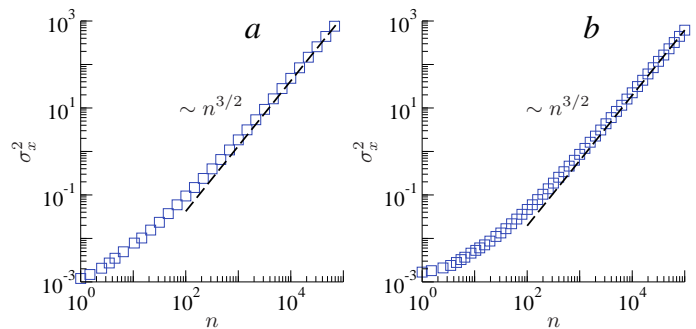}}
  \caption{(color online) Monte Carlo simulations results for the
    variance $\sigma_x^2$ of the BI displacement in stripes with $L_x
    = 10^4$ and $L_y = 3$ (panel {\it a}) and rectangular capillaries
    with $L_x = 10^4$ and $L_y = L_z =3$ (panel {\it b}) for $\rho_0 =
    0.002$ and $\beta F=100$. The dashed curves in both panels have
    the slope $n^{3/2}$. }
  \label{fig-7}
\end{figure}

The following  remark is  in order: An  appearance of  the logarithmic
factor   can  be   interpreted   as   a  sign   that   $d=2$  is   the
\textit{marginal} dimension for this system, above which (e.g., in 3D)
one  should expect  a usual  diffusive growth  of  $\sigma_x^2$ (with,
however,  a larger  prefactor than  in the  perpendicular to  the bias
directions),  while in  1D,  i.e., for  a  single-file diffusion,  the
effect  should be stronger  and $\sigma_x^2$  would get  an additional
power-law dependence on  time.  For 3D, indeed, such  a scenario seems
plausible.   We  have  numerically   simulated  a  BI  dynamics  in  a
single-file symmetric lattice gas and realized that, surprisingly, the
growth law of $\sigma_x^2$ is \textit{not affected} by the bias, i.e.,
$\sigma_x^2 \sim  n^{1/2}$. Moreover,  our simulations show  that even
the prefactor in this dependence is \textit{independent} of $\beta F$.
On  contrary,  our  preliminary  estimates  suggest  that  a  strongly
superdiffusive behaviour $\sigma_x^2 \sim  n^{3/2}$ will take place in
quasi-1D systems  - infinite  two-dimensional stripes ($L_x  = \infty$
and finite $L_y$)  and three-dimensional rectangular capillaries ($L_x
= \infty$  and finite  $L_y$ and $L_z$).   Indeed, our  Eq. (\ref{11})
tells that $\sigma_x^2 \sim n/\chi_n$.  By definition, $\chi_n = S_n -
S_{n - 1}$,  where $S_n$ is the mean number  of distinct sites visited
by any individual vacancy up to time $n$. For infinite stripes one has
$S_n \sim n^{1/2}/L_y$, while  $S_n \sim n^{1/2}/L_y L_z$ for infinite
rectangular capillaries. Therefore, in  both cases we have $\sigma_x^2
\sim n^{3/2}$ which strongly superdiffusive behaviour is unambiguously
confirmed by our numerical simulations (see, Fig.~\ref{fig-7}).

We  note finally  that a  similar superdiffusive  behavior $\sigma_x^2
\sim  n^{\zeta}$  with  $\zeta  \approx  1.45$ has  been  observed  in
Molecular  Dynamics simulations  of a  biased intruder  dynamics  in a
glass-forming binary Yukawa liquid in \cite{binder} and more recently,
in  \cite{heuer}.  Taken together,  our  results  and  the results  of
Refs.\cite{binder,heuer} hint  at a possibility of  encountering a new
physical  phenomenon -  a field-induced  superdiffusive  broadening of
fluctuations in  dynamics of a biased intruder  in quiescent molecular
crowding  environments. This  is rather  surprising, since  such dense
environments are commonly associated with a subdiffusive behavior.

\acknowledgments  We thank  K.   Binder, S.   N.   Majumdar, A.   Law,
F. Ritort, P.  Royall, M. Schick,  U. Seifert and M. Weiss for helpful
discussions.   OB acknowledges  support  from the  ERC starting  Grant
FPTOpt-277998. CMM  is supported by the European  Research Council and
the Academy of Finland.  Partial support from the ESF Research Network
"Exploring the Physics of Small Devices" is also acknowledged.

%%%%%%%%%%
%%%%%%%%%%

\end{document}